\documentclass[conference, 10pt]{IEEEtran}
\IEEEoverridecommandlockouts
\usepackage{cite}
\usepackage{amsmath,amssymb,amsfonts}
\usepackage{graphicx}
\usepackage{textcomp}
\usepackage{amsmath}
\usepackage{booktabs} 
\usepackage{colortbl}
\usepackage{multirow}
\usepackage{stfloats}
\usepackage{hyperref}       
\usepackage{url}            
\usepackage{booktabs}       
\usepackage{amsfonts}       
\usepackage{nicefrac}       
\usepackage{microtype}      
\usepackage{xcolor}         
\usepackage{amsmath}
\usepackage{tikz}
\usepackage{braket} 

\usepackage{quantikz}
\usepackage{subcaption}
\usepackage{graphicx}   
\usepackage{adjustbox}  

\newcommand*\circled[1]{\tikz[baseline=(char.base)]{
            \node[shape=circle,draw,inner sep=2pt] (char) {#1};}}
\usepackage{orcidlink}
\usepackage{comment}
\usepackage{stfloats}
\usepackage{setspace}
\usepackage{amssymb}
\usepackage{braket}
\usepackage{xcolor}
\usepackage{svg}
\usepackage{pifont}
\newcommand{\cmark}{\ding{51}}  
\newcommand{\xmark}{\ding{55}}  

\usepackage[compact]{titlesec}
\usepackage[font=footnotesize,skip=0pt]{caption}

\titlespacing\section{0pt}{0.3\baselineskip}{0.2\baselineskip}
\titlespacing\subsection{0pt}{0.2\baselineskip}{0.1\baselineskip}
\titlespacing\subsubsection{0pt}{0.15\baselineskip}{0.1\baselineskip}

\usepackage{algpseudocode} 
\usepackage[most]{tcolorbox}
\tcbuselibrary{skins, breakable, hooks}

\newtcolorbox{algobox}[1][]{
  enhanced, breakable,
  colback=white, colframe=black,
  boxrule=0.6pt, arc=2pt,
  left=8pt, right=8pt, top=8pt, bottom=8pt,
  attach boxed title to top left={yshift=-2pt, xshift=6pt},
  boxed title style={
    colback=black!5,   
    colframe=black,
    boxrule=0.6pt,
    arc=2pt
  },
  coltitle=black,       
  title={\textbf{#1}}
}
\newcommand{\rpoint}[1]{\scalebox{0.85}{\circled{{\fontfamily{pcr}\selectfont\footnotesize #1}}}}

\usepackage{enumitem}

\setlist[itemize]{leftmargin=*}%
\setlist[enumerate]{leftmargin=*}%

\def\BibTeX{{\rm B\kern-.05em{\sc i\kern-.025em b}\kern-.08em
    T\kern-.1667em\lower.7ex\hbox{E}\kern-.125emX}}

\usepackage[compact]{titlesec}

\titlespacing\section{0pt}{0.3\baselineskip}{0.2\baselineskip}
\titlespacing\subsection{0pt}{0.2\baselineskip}{0.1\baselineskip}
\titlespacing\subsubsection{0pt}{0.15\baselineskip}{0.1\baselineskip}

\begin{document}

\title{Beyond Logical Circuits: Hardware-Aware Analysis of Expressibility and Trainability in Variational Quantum Algorithms\\
}

\author{\IEEEauthorblockN{Muhammad Kashif\orcidlink{0000-0003-2023-6371}\textsuperscript{1,2}, and Muhammad Shafique\orcidlink{0000-0002-2607-8135}\textsuperscript{1,2}\\
 \IEEEauthorblockA{
 \textsuperscript{1}eBRAIN Lab, Division of Engineering, New York University Abu Dhabi (NYUAD), Abu Dhabi, UAE\\
 \textsuperscript{2}Center for Quantum and Topological Systems (CQTS), NYUAD Research Institute, NYUAD, Abu Dhabi, UAE\\
 \{muhammadkashif, muhammad.shafique\}@nyu.edu
}}
\vspace{-20pt}
}

\maketitle
\begin{abstract}
Variational quantum algorithms (VQAs) rely on parameterized quantum circuits (PQCs), whose performance is governed by expressibility and trainability. Existing studies typically evaluate these properties at the logical circuit level, implicitly assuming that designed PQCs remain unchanged during hardware execution. In practice, however, hardware-aware transpilation modifies circuit structure through qubit mapping, routing, and basis decomposition, potentially altering PQC behavior.
In this paper, we perform a systematic hardware-aware analysis of expressibility and trainability by comparing logical and transpiled PQCs across multiple ansatz families, qubit counts, and circuit depths. Expressibility is measured using fidelity-based KL divergence, while trainability is quantified through gradient variance.
Our results show that transpilation acts as an implicit architectural perturbation, producing strongly ansatz-dependent effects. Expressibility deviations exceed upto 125\% in some cases, while trainability variations reach up to 25\%. Structured ansatzes are generally more robust, whereas highly entangled architectures are more sensitive to transpilation-induced transformations. We further show that transpilation can alter the commonly assumed expressibility-trainability trade-off, demonstrating that logical-level analyses may not reliably predict hardware-level behavior. These findings highlight the importance of hardware-aware evaluation for accurate characterization of VQAs.

\end{abstract}

\begin{IEEEkeywords}
Variational quantum algorithms, Quantum hardware, NISQ, Expressibility, Trainability,
Quantum Transpilation, Parameterzied quantum circuits 
\end{IEEEkeywords}

\begin{spacing}{1.0}
\section{Introduction} \label{sec:intro}

In the noisy intermediate-scale quantum (NISQ) era, variational quantum algorithms (VQAs) have emerged as a leading framework for practical quantum computing applications~\cite{kashif_PP}. VQAs combine parameterized quantum circuits (PQCs) with classical optimization loops and have been widely applied to optimization, quantum chemistry, and machine learning~\cite{Cerezo_2021,kashif2021design,innan2025quav}. The performance of VQAs is significantly dependent on the underlying PQCs, making their design a critical factor in determining the effectiveness and scalability of VQAs~\cite{kashif2025computational}.
Two key properties govern PQC performance: \textit{expressibility} and \textit{trainability}. Expressibility measures how well a PQC approximates Haar-random quantum states or unitaries, typically quantified through distributional divergence metrics~\cite{Sim_2019,liu2025analysis,roseler2026find,walid_2025}. Trainability refers to the ease of optimizing PQC parameters and is commonly analyzed through optimization landscapes and gradient statistics~\cite{shao2025diagnosing,Kashif_2024_resqnets}. Poor trainability often manifests as barren plateaus, where gradients vanish exponentially with system size, making optimization increasingly difficult~\cite{mcclean2018barren,kashif2024alleviating,atallah2025investigating,kashif2023impact}.
These properties are closely related and often exhibit a trade-off: highly expressive circuits, particularly those approaching unitary 2-designs, are more prone to barren plateaus, whereas more structured circuits remain trainable at the expense of reduced expressive power~\cite{Holmes_2022,kashif2023unified,Cerezo_2021_CF}. Although hardware noise and cost-function locality can also induce barren plateaus independently~\cite{wang2024entanglement,Cerezo_2021_CF,ahmed2025comparative}, this inverse relationship generally holds in noiseless settings. Consequently, diverse ansatz families, including hardware-efficient, tensor-network-inspired, and problem-specific circuits, have been proposed to balance expressibility and trainability~\cite{Kandala_2017,park2024HEA,berezutskii2025tensor,sugawara2025ttn,wang2025variational}.

Most existing studies evaluate these properties at the logical circuit level, implicitly assuming that the designed PQC is identical to the circuit executed on hardware. In practice, however, PQCs must undergo hardware-aware compilation (transpilation)\footnote{Transpilation and hardware-aware compilation are used interchangeably in this paper} to satisfy constraints such as limited qubit connectivity, native gate sets, and device topology~\cite{qiskit_transpilation,Louamri_2024}. Transpilation can substantially alter circuit structure by increasing depth, inserting additional two-qubit gates, and modifying entanglement patterns~\cite{kashif2026late}. As a result, the hardware-executed circuit may differ significantly from the intended logical design, not only in resource overhead but also in its fundamental properties.
While prior work has extensively studied the impact of transpilation on resource costs, noise sensitivity, and reliability~\cite{stefano2024empirical,dilillo2023understanding,kashif2025faqnas,huo2025revisiting,roy2025forensics,kashif2026closing}, its effect on expressibility and trainability remains largely unexplored. If transpilation changes the expressive power or optimization landscape of PQCs, then logical-level analyses may not accurately predict hardware-level behavior. This motivates the need for hardware-aware evaluation frameworks that explicitly account for transpilation-induced transformations.

\subsection{Motivational Analysis} \label{sec:mot_analysis}
\begin{figure}[h]
    \centering
    \includegraphics[width=0.98\linewidth]{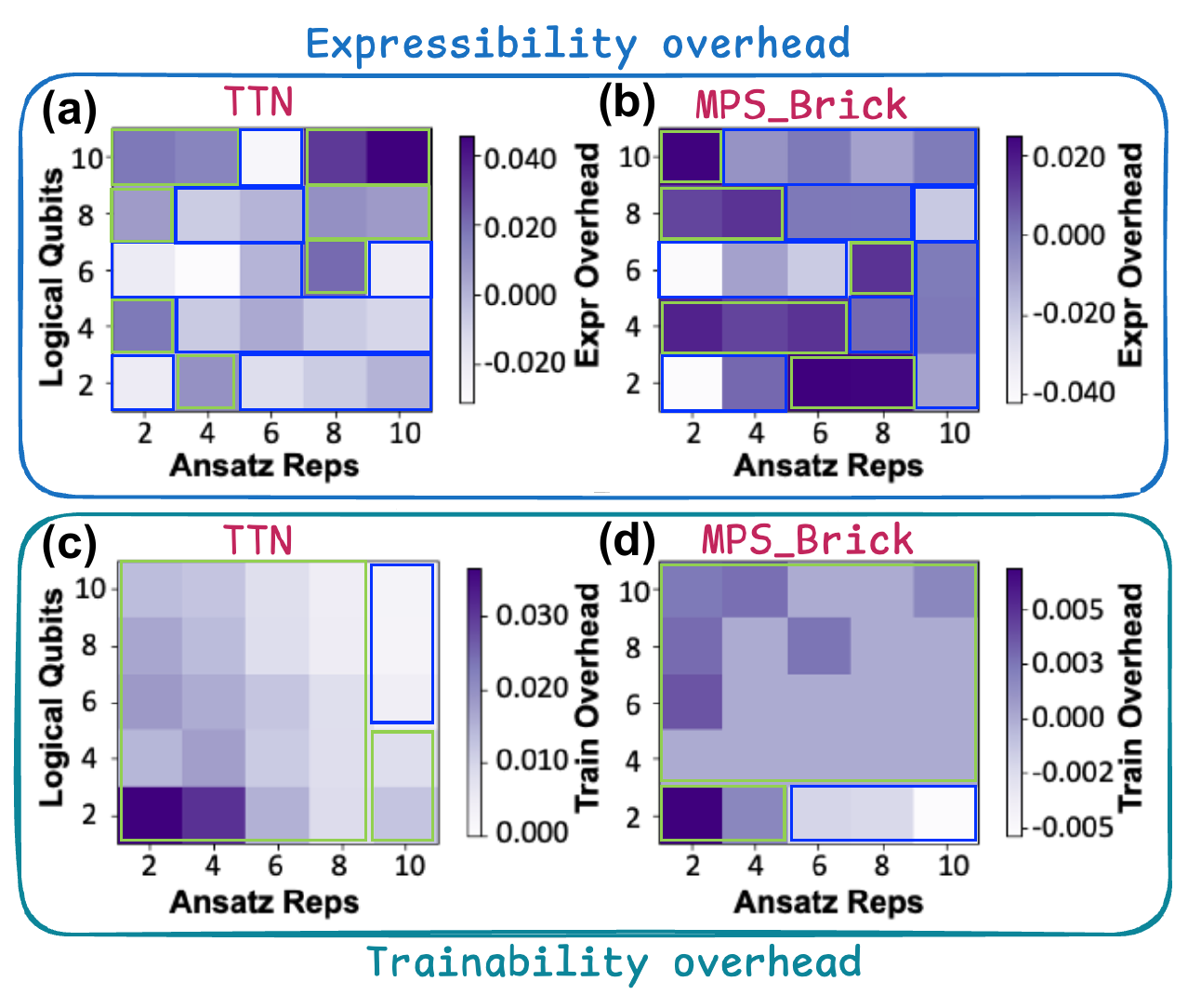}
    \caption{Transpilation-induced overhead in expressibility and trainability for the \textit{TTN} and \textit{MPS\_Brick} ansatz families across different logical qubit counts and ansatz repetitions. (a) and (b) show the expressibility overhead, where positive values (green boxes) indicate reduced expressibility after transpilation and negative values (blue boxes) indicate improvement or no change. (c) and (d) show the trainability overhead, where positive values correspond to improved trainability after transpilation, while blue-box regions indicate preserved or reduced trainability after transpilation.}
    \label{fig:motivational_study}
\end{figure}

Fig.~\ref{fig:motivational_study} illustrates the impact of transpilation on the expressibility and trainability for two ansatz families: the tensor tree network (\textit{TTN})~\cite{sugawara2025ttn} and matrix product state \textit{MPS\_Brick}~\cite{fan2023quantum}. Results are reported for different qubit counts and ansatz repetitions using transpiler optimization level $3$ with the \textit{SABRE} layout~\cite{zou2024lightsabre,qiskit_transpilation_sabre}. The transpilation overhead is defined as the difference between metrics evaluated on transpiled and logical circuits.
Expressibility overhead is measured using KL divergence:
\begin{equation}\label{eq:expr_overhead} \Delta \mathcal{E}_{\mathrm{KL}} = \mathcal{E}_{\mathrm{KL}}^{\mathrm{transpiled}} - \mathcal{E}_{\mathrm{KL}}^{\mathrm{logical}} \end{equation}

where $\Delta \mathcal{E}_{\mathrm{KL}} > 0$ indicates reduced expressibility after transpilation, while negative values indicate improvement. Trainability overhead is measured using gradient variance:

\begin{equation} \label{eq:train_overhead} \Delta \mathrm{GradVar} = \mathrm{GradVar}_{\mathrm{transpiled}} - \mathrm{GradVar}_{\mathrm{logical}} \end{equation}

where positive values correspond to improved trainability after transpilation.
For expressibility overhead, green-box regions indicate configurations where transpilation increases the KL divergence relative to the logical circuit, resulting in reduced expressibility, whereas blue-box regions correspond to reduced or near-zero KL divergence, indicating preserved or improved expressibility.
The \textit{TTN} ansatz in Fig.~\ref{fig:motivational_study}(a) exhibits predominantly positive overhead at higher qubit counts and ansatz repetitions, showing that transpilation generally reduces its expressibility in deeper configurations. In contrast, \textit{MPS\_Brick} in Fig.~\ref{fig:motivational_study}(b) presents alternating green and blue regions, indicating that transpilation can either improve or degrade expressibility depending on the circuit configuration.

For trainability overhead, green-box regions correspond to positive gradient variance overhead, indicating improved trainability after transpilation, whereas blue-box regions indicate near-zero or negative overhead, corresponding to preserved or slight degradation in trainability.
The \textit{TTN} ansatz in Fig.~\ref{fig:motivational_study}(c) shows strong positive overhead at lower ansatz repetitions, suggesting that transpilation significantly improves trainability in shallow circuits, while the highlighted blue region at larger qubit count indicates diminishing impact. For \textit{MPS\_Brick} ansatz in Fig.~\ref{fig:motivational_study}(d), the overhead is mostly mostly close to zero, with only localized positive regions, implying that transpilation has comparatively minor effects on its trainability.

Overall, these results demonstrate that transpilation can substantially alter both expressibility and trainability in an ansatz-dependent manner. Therefore, logical-level evaluations alone may not accurately reflect hardware-level behavior, motivating hardware-aware analyses that explicitly account for transpilation-induced effects.

\begin{table*}[h]
\centering
\caption{State-of-the-art comparison of VQAs performance w.r.t expressibility, trainability, and transpilation effects. To the best of our knowledge, no prior work has conducted a systematic, metric-driven comparison of expressibility and trainability between logical circuits and their transpiled counterparts across multiple ansatz families.}
\label{tab:related_work}
\begin{adjustbox}{max width=.99\linewidth}
\begin{tabular}{l||c|c|c|c|c|c}
\hline
\rowcolor{gray!20}
\textbf{Work} 
& \textbf{Expressibility} 
& \textbf{Trainability} 
& \textbf{Multiple Ansatzes} 
& \textbf{Transpilation-Aware} 
& \textbf{Logical vs Physical} 
& \textbf{Depth/Qubit Sweeps} \\
\hline

\cite{Sim_2019} 
& \cmark & \xmark & \cmark & \xmark & \xmark & \cmark \\
\hline

\rowcolor{gray!10}
\cite{walid_2025} 
& \cmark & \cmark & \cmark & \xmark & \xmark & \xmark \\
\hline

\rowcolor{gray!5}
\cite{mcclean2018barren} 
& \xmark & \cmark & \xmark & \xmark & \xmark & \cmark \\
\hline

\rowcolor{gray!10}
\cite{kashif2024alleviating} 
& \xmark & \cmark & \xmark & \xmark & \xmark & \cmark \\
\hline

\rowcolor{gray!5}
\cite{heyraud2023efficient} 
& \xmark & \cmark & \xmark & \xmark & \xmark & \cmark \\
\hline

\rowcolor{gray!10}
\cite{Holmes_2022} 
& \cmark & \cmark & \cmark & \xmark & \xmark & \cmark \\
\hline

\rowcolor{gray!5}
\cite{Cerezo_2021_CF} 
& \xmark & \cmark & \xmark & \xmark & \xmark & \cmark \\
\hline

\rowcolor{gray!10}
\cite{stefano2024empirical} 
& \xmark & \xmark & \cmark & \cmark & \cmark & \xmark \\
\hline

\rowcolor{green!15}
\textbf{Ours} 
& \cmark & \cmark & \cmark & \cmark & \cmark & \cmark \\
\hline

\end{tabular}
\end{adjustbox}
\end{table*}

\subsection{Our Novel Contributions}

\begin{itemize}

\item \textbf{Limitations of logical-level evaluations:}
We first demonstrate that logical-level expressibility and trainability metrics are not always reliable predictors of hardware-level behavior. PQCs can exhibit substantially different characteristics after hardware-aware compilation, highlighting the need for transpilation-aware evaluation of VQAs on NISQ devices (\textbf{Section~\ref{sec:mot_analysis}})

\item \textbf{Hardware-aware evaluation of parameterized quantum circuits:}
We present the first systematic study of PQC expressibility and trainability at both the \textit{logical} and \textit{transpiled} levels. By comparing circuits before and after hardware-aware compilation, we establish a quantitative framework for analyzing transpilation-induced effects beyond conventional resource metrics such as depth and gate count (\textbf{Section~\ref{sec:methodology}}).

\item \textbf{Systematic multi-ansatz benchmarking:}
We conduct a comprehensive evaluation across multiple ansatz families, qubit counts, and circuit depths, enabling large-scale analysis of how transpilation affects PQC functional properties across diverse architectures (\textbf{Section~\ref{sec:methodology}}).

\item \textbf{Transpilation-induced changes in expressibility and trainability:}
Our results show that hardware-aware compilation can significantly and non-monotonically reshape both expressibility and trainability. In particular, transpiler optimization strategies can introduce abrupt depth-dependent changes in gradient statistics, indicating that transpiler heuristics can qualitatively alter VQA optimization landscapes (\textbf{Section~\ref{sec:results}}).

\item \textbf{Ansatz-dependent robustness to transpilation:}
We identify strong architecture-dependent sensitivity to compilation. Structured ansatzes such as matrix-product-state (MPS) and tree tensor network (\textit{TTN}) circuits exhibit different robustness profiles compared to hardware-efficient and fully connected architectures, highlighting the importance of hardware-aware PQC design (\textbf{Section~\ref{sec:results}}).

\item \textbf{Breakdown of the expressibility--trainability trade-off under transpilation.}
We demonstrate that the commonly assumed expressibility--trainability trade-off does not uniformly hold after hardware-aware compilation. Transpilation can modify these properties in a decoupled and ansatz-dependent manner, where changes in expressibility are not necessarily reflected in trainability (\textbf{Section~\ref{sec:results}}).

\end{itemize}

In summary, we argue that analyses of VQAs conducted at the logical design level characterize only their algorithmic capacity. This is analogous to studying neural networks without accounting for deployment factors such as quantization, memory layout, or hardware scheduling, where theoretical insights may not fully translate upon real-world deployment. Our results demonstrate that hardware-aware compilation is not merely a passive preprocessing step, but an active transformation that can fundamentally alter the representational capacity and optimization landscape of VQAs.


\section{Related Work}
The theoretical properties of PQCs have been extensively studied from multiple perspectives. The optimization difficulty of VQAs is now well understood to be closely tied to the barren plateau phenomenon, first identified in~\cite{mcclean2018barren}, and subsequently generalized to account for cost-function locality, circuit depth, entanglement structure and random parameter initialization~\cite{Marrero:2020,Cerezo_2021_CF,kashif:2024dilemma,Holmes_2022,kashif2024alleviating}.
In parallel, the representational capacity of PQCs has been formalized through the notion of expressibility, most notably via fidelity-distribution-based measures and related design metrics~\cite{nakaji:2021,ragone:2022,zhang2025learning}.

A growing body of work has also investigated the interplay between expressibility and trainability, showing that highly expressive circuits often exhibit poor gradient scaling, while structured ansatzes can mitigate barren plateaus at the cost of reduced solution space~\cite{Sim_2019,Holmes_2022}.
Motivated by these insights, a wide variety of ansatz families have been proposed, ranging from hardware-efficient designs~\cite{park2024HEA,wang2024entanglement} to tensor-network-inspired architectures such as matrix product states~\cite{fan2023quantum}, tree tensor networks~\cite{lazzarin2022multi,sahoo2022quantum}. Qiskit-native templates such as \textit{TwoLocalRYRZ}~\cite{qiskit_ansatz_2_local}, \textit{EfficientSU2}~\cite{qiskit_ansatz_eff_su2}, and \textit{RealAmplitudes}~\cite{qiskit_ansatz_real_amp_ansatz} have further popularized systematic explorations of entanglement patterns and parameterizations in practical workflows. These architectures have been benchmarked primarily at logical level, under idealized assumptions about circuit execution on the actual hardware.

Quantum circuit transpilation has also been widely studied, with a focus on fault-propagation, resource overhead, qubit mapping, SWAP routing, and gate-level optimization under hardware constraints~\cite{dilillo2023understanding,younis2022}. However, majority of existing works in this regard treat transpilation primarily as a resource optimization, and do not consider expressibility and trainability analysis which is mostly studied at the logical circuit level. A summary of recent works is provided in Table~\ref{tab:related_work}. 


\section{Background and Preliminaries}\label{sec:background}

\subsection{Variational Quantum Algorithms}
\label{sec:background_PQC}

\begin{figure}[h]
    \centering
    \includegraphics[width=1.0\linewidth]{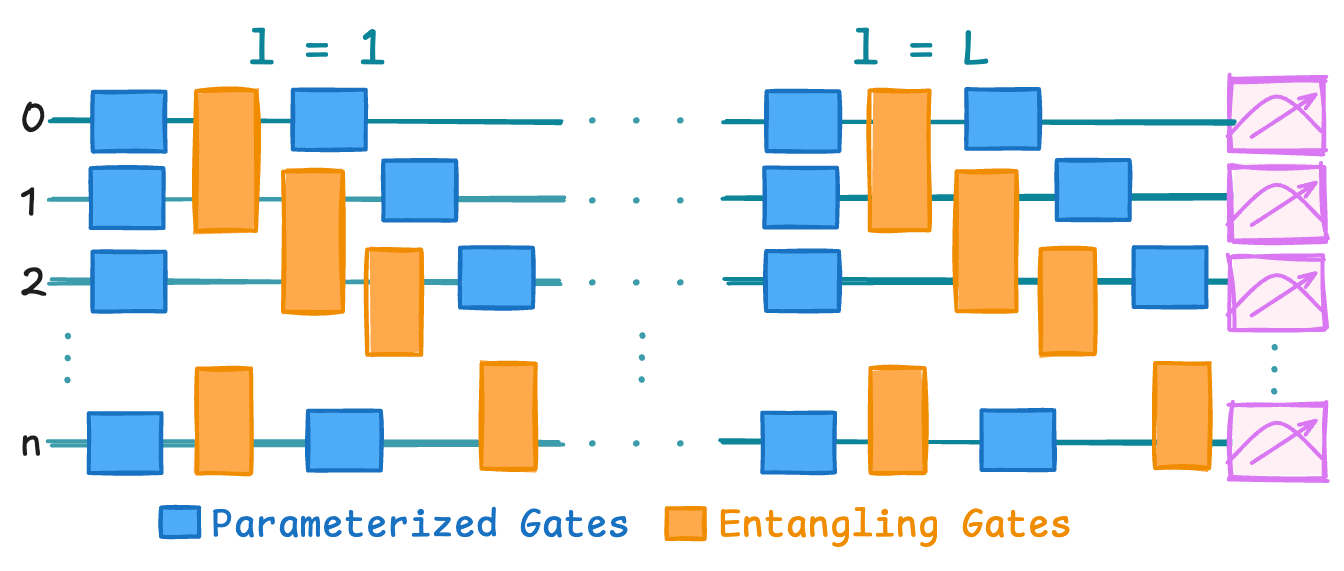}
    \caption{An illustration of PQC used in variational quantum algorithms}
    \label{fig:PQC}
\end{figure}

VQAs are based on the optimization of PQCs within a hybrid quantum--classical feedback loop. A PQC consists of a sequence of quantum gates whose action depends on a set of parameters \( \boldsymbol{\theta} = (\theta_1, \dots, \theta_P) \). Starting from a fixed reference state, typically the computational basis state \( \lvert 0 \rangle^{\otimes n} \), the circuit prepares a parameterized quantum state:
\begin{equation}\label{eq:log_state}
\lvert \psi(\boldsymbol{\theta}) \rangle = U(\boldsymbol{\theta}) \lvert 0 \rangle^{\otimes n}
\end{equation}
where $n$ is number of qubits and \( U(\boldsymbol{\theta}) \) is a unitary operator realized by a finite-depth quantum circuit composed of parameterized single-qubit rotations and  multi-qubit entangling gates, as depicted in Fig.~\ref{fig:PQC}.
The parameters \( \boldsymbol{\theta} \) are optimized to minimize a cost function \( C(\boldsymbol{\theta}) \) defined as the expectation value of an observable \( O \):
\begin{equation}
C(\boldsymbol{\theta}) = \langle \psi(\boldsymbol{\theta}) \rvert O \lvert \psi(\boldsymbol{\theta}) \rangle
= \langle 0 \rvert U^\dagger(\boldsymbol{\theta}) \, O \, U(\boldsymbol{\theta}) \lvert 0 \rangle.
\label{eq:cost_function}
\end{equation}
The optimization is performed using an optimizer, which iteratively updates the parameters according to estimates of the gradient \( \nabla_{\boldsymbol{\theta}} C(\boldsymbol{\theta}) \). 
A common and widely studied class of PQCs is obtained by composing \( L \) repetitions (layers) of a fixed parameterized block:
\begin{equation}\label{eq:PQC}
U(\boldsymbol{\theta}) = \prod_{\ell=1}^{L} U_\ell(\boldsymbol{\theta}_\ell),
\end{equation}
where each layer \( U_\ell(\boldsymbol{\theta}_\ell) \) typically consists of a layer of single-qubit rotations followed by an entangling gate pattern. 
Different choices of the layered block structure and entangling patterns define different \emph{ansatz families}, which impose different inductive biases on the variational model. 

%
\subsection{Quantum Compilation and Transpilation}
\label{sec:background_transpilation}

\begin{figure}[h]
    \centering
    \includegraphics[width=1.0\linewidth]{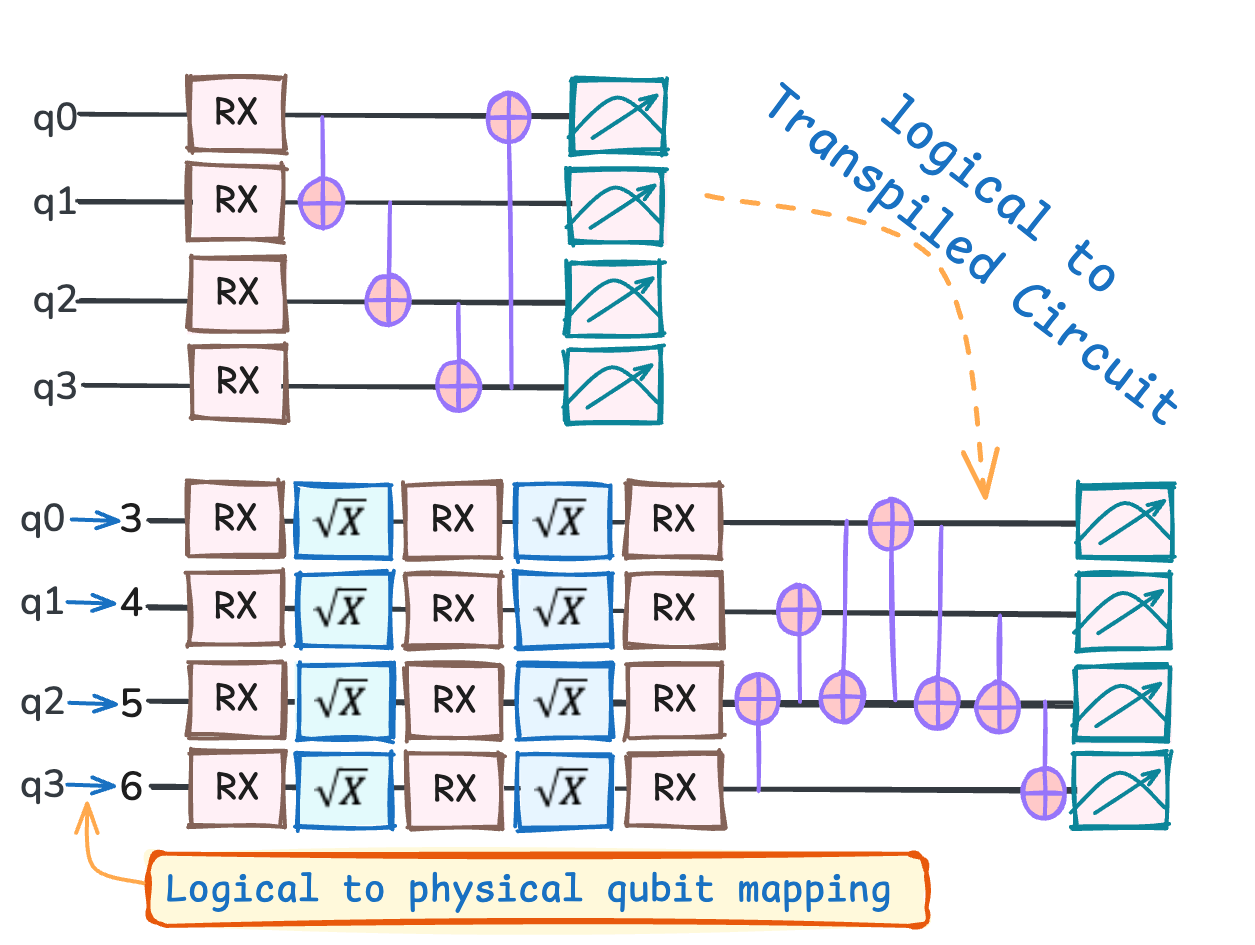}
    \caption{Illustration of quantum circuit transpilation on IBM FakeWashington backend. For a small $4$-qubit circuit designed at logical level, the gate count and depth is increased significantly.}
    \label{fig:log_phy_ckt}
\end{figure}

PQCs are typically defined at the \emph{logical level}, which defines a unitary \( U_{\mathrm{logical}}(\boldsymbol{\theta}) \) on \( n \) qubits, and represents the intended behavior desired by the user. It assumes ideal qubits, full connectivity, and a universal gate set. 
However, to execute the PQCs on real hardware, they must be \emph{transpiled} into a hardware-compatible form that respects device connectivity and native gate constraints.
Transpilation involves three main steps~\cite{zou2024lightsabre,qiskit_transpilation_sabre}: \emph{(i)~layout selection}, which maps logical qubits to physical qubits; \emph{(ii)~routing}, where SWAP or two-qubit gates are inserted to satisfy connectivity constraints,  and \emph{(iii)~basis decomposition and optimization}, where gates are expressed in the native gate set. 
The \emph{transpiled circuit} implements \( U_{\mathrm{transpiled}}(\boldsymbol{\theta}) \) on a (possibly larger) set of physical qubits: 
\begin{equation}
U_{\mathrm{transpiled}}(\boldsymbol{\theta}) = W^\dagger \, U_{\mathrm{logical}}(\boldsymbol{\theta}) \, W
\label{eq:compilation_equivalence}
\end{equation}
where \( W \) captures the effect of layout, routing, and gate synthesis.
Transpilation can significantly alter the \emph{structure} of the circuit, including depth, gate composition, entanglement pattern, and sometimes even the number of active qubits. As a result, the implemented circuit \( U_{\mathrm{transpiled}}(\boldsymbol{\theta}) \) may exhibit different properties than its logical counterpart. An illustration of transpilation effects is shown in Fig.~\ref{fig:log_phy_ckt}). Due to significant structural changes with additional single- and two-qubit gates, transpilation should not be viewed only as a resource-level transformation, but as a process that can potentially modify their key properties of PQCs.
\vspace{-15pt}
%
%
%
\subsection{Expressibility of PQCs}
\label{sec:background_expressibility}

The expressibility of a PQC quantifies how well the states it generates approximate the uniform (Haar) distribution over the Hilbert space~\cite{Sim_2019,ragone:2022,Holmes_2022}. Intuitively, a highly expressive ansatz is capable of exploring a large portion of the Hilbert space, while a poorly expressive ansatz generates states confined to a restricted subspace.
A widely used approach to characterize expressibility is based on the distribution of pairwise state fidelities~\cite{Sim_2019}. For two independently sampled parameter vectors \( \boldsymbol{\theta} \) and \( \boldsymbol{\theta}' \), the corresponding output states are:
%
\begin{equation}
\lvert \psi(\boldsymbol{\theta}) \rangle = U(\boldsymbol{\theta}) \lvert 0 \rangle^{\otimes n} 
\quad, \quad
\lvert \psi(\boldsymbol{\theta}') \rangle = U(\boldsymbol{\theta}') \lvert 0 \rangle^{\otimes n}
\end{equation}

and their fidelity is defined as:
\begin{equation}\label{eq:fidelity}
F(\boldsymbol{\theta}, \boldsymbol{\theta}') = \left| \langle \psi(\boldsymbol{\theta}) \mid \psi(\boldsymbol{\theta}') \rangle \right|^2
\end{equation}

The distribution of these fidelities can be compared to the analytical fidelity distribution induced by Haar-random states in a Hilbert space of dimension \( d = 2^n \):
\begin{equation}\label{eq:haar}
P_{\mathrm{Haar}}(F) = (d - 1)(1 - F)^{d - 2}
\end{equation}

which corresponds to a Beta$(1, d - 1)$ distribution.
The discrepancy between the circuit-induced distribution and the Haar distribution is commonly quantified using the Kullback--Leibler (KL) divergence:
\begin{equation}\label{eq:KLD}
D_{\mathrm{KL}} \big( P_{\mathrm{ansatz}} \,\|\, P_{\mathrm{Haar}} \big)
= \int P_{\mathrm{circuit}}(F)\, \log \frac{P_{\mathrm{circuit}}(F)}{P_{\mathrm{Haar}}(F)} \, dF
\end{equation}

Lower values of KL divergence indicate that the circuit-induced distribution is closer to the Haar distribution, and hence correspond to higher expressibility, whereas larger values indicate reduced expressibility.

\begin{figure*}
    \centering
    \includegraphics[width=1.0\linewidth]{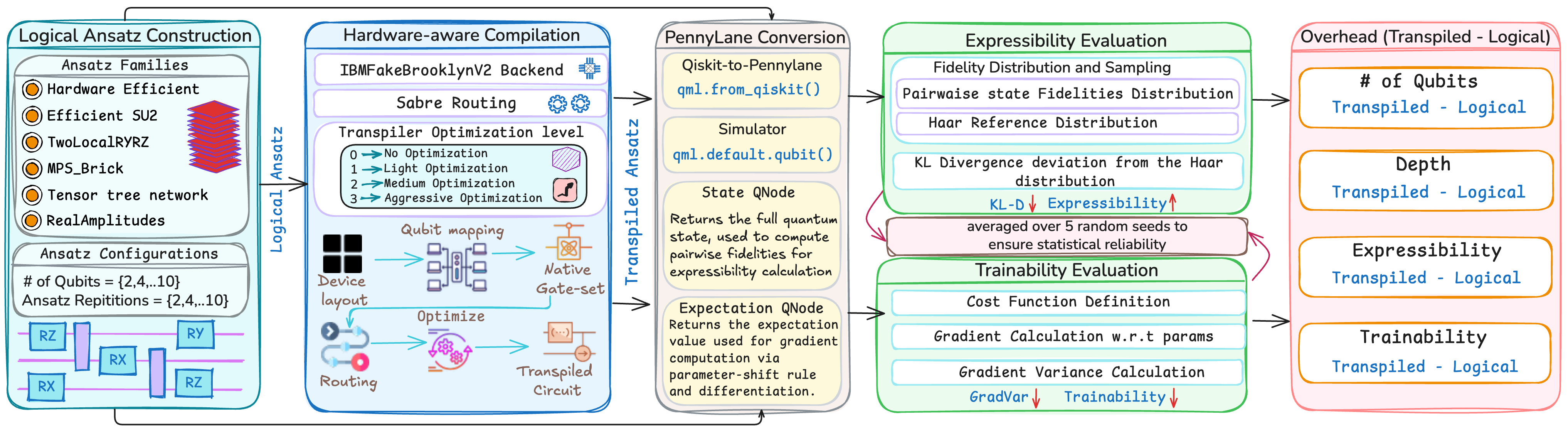}
    \caption{Overview of the our methodology. Logical PQCs from multiple ansatz families are first constructed across varying qubit counts and depths. These circuits are then transpiled using a hardware-aware compilation pipeline (IBM FakeBrooklynV2 backend with SABRE routing and varying optimization levels). The resulting logical and transpiled circuits are converted to PennyLane for unified statevector simulation. Expressibility is evaluated via fidelity distributions and KL divergence, while trainability is quantified using gradient variance. Finally, overhead is computed as the difference between transpiled and logical metrics, capturing transpilation-induced changes in circuit structure and functional properties.}
    \label{fig:method_log_vs_phys}
\end{figure*}

\subsection{Trainability of PQCs}
\label{sec:background_trainability}

The trainability of a PQC refers to the ability to efficiently optimize its parameters. A central challenge in VQAs is the \emph{barren plateau} phenomenon, where gradients vanish exponentially with system size, making optimization increasingly difficult~\cite{mcclean2018barren}.
A widely used metric for trainability is the variance of the gradients of the cost function with respect to circuit parameters. Consider a cost function of the form:
\begin{equation}\label{eq:Cost_F_trainability}
C(\boldsymbol{\theta}) = \langle \psi(\boldsymbol{\theta}) \rvert O \lvert \psi(\boldsymbol{\theta}) \rangle
\end{equation}
where \( O \) is a Hermitian observable and \( |\psi(\boldsymbol{\theta})\rangle = U(\boldsymbol{\theta}) |0\rangle^{\otimes n} \). The gradient with respect to a parameter \( \theta_i \) is given by:
\begin{equation}
g_i(\boldsymbol{\theta}) = \frac{\partial C(\boldsymbol{\theta})}{\partial \theta_i}
\end{equation}

Trainability is then quantified via the variance of gradients over randomly sampled parameters:
\begin{equation}\label{eq:grad_variance_def}
\mathrm{Var}[g_i] = \mathbb{E}_{\boldsymbol{\theta}} \left[ g_i(\boldsymbol{\theta})^2 \right]
- \left( \mathbb{E}_{\boldsymbol{\theta}} \left[ g_i(\boldsymbol{\theta}) \right] \right)^2
\end{equation}

In barren plateau regimes, this variance decays exponentially with system size, leading to poor trainability. Larger gradient variance indicates stronger gradient signals and improved trainability.


\section{Our Methodology}
\label{sec:methodology}

In this paper, we investigate whether hardware-aware compilation affects PQCs beyond just resource overhead. Specifically, we analyze how transpilation alters the expressibility and trainability of variational ansatzes.
To isolate these effects, we compare logical circuits (designer's intent) with their transpiled counterparts (hardware-executable circuits) while keeping the parameter count fixed. Both circuit representations are evaluated using the same noiseless statevector simulator, ensuring that observed differences arise purely from transpilation-induced structural transformations.
We consider six ansatz families across different qubit counts, circuit depths, and transpiler optimization levels. Fig.~\ref{fig:method_log_vs_phys} summarizes the overall methodology.

\begin{figure*}
    \centering
    \includegraphics[width=1.0\linewidth]{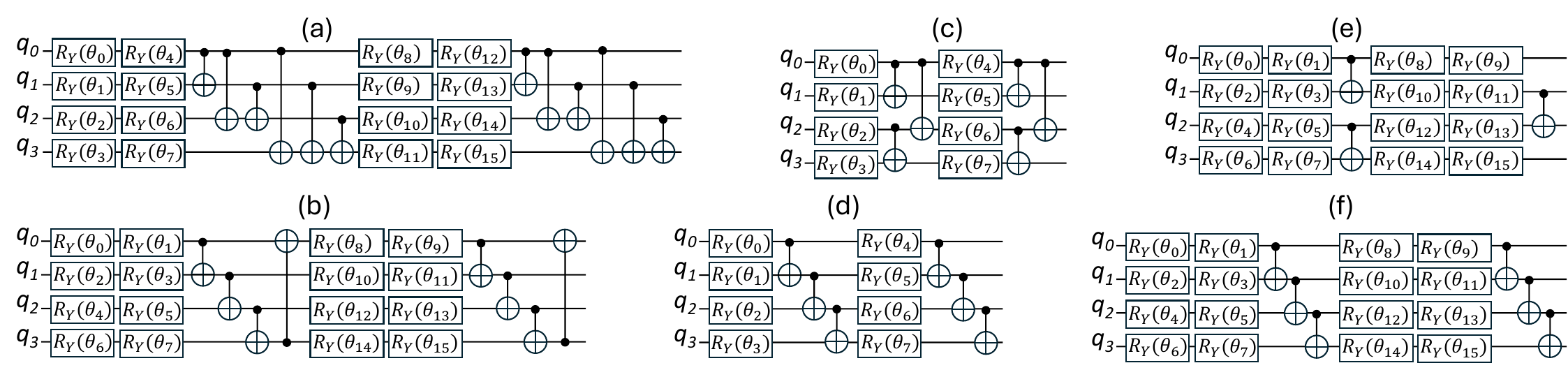}
    \caption{Ansatz Families used in this paper. (a) \textit{EfficientSU2} Ansatz, (b) HEA$\_$Ring Ansatz, (c) \textit{TTN}\_Tree ansatz, (d) Real Amplitudes, (e) \textit{MPS\_Brick}, (f) \textit{TwoLocalRYRZ}}
    \label{fig:ansatzes}
\end{figure*}

\subsection{Logical Ansatz Construction}
\label{sec:method_ansatz}

We consider a set of PQCs belonging to six representative ansatz families: hardware-efficient with ring entanglement (\textit{\textit{HEA\_Ring}}), matrix-product-state (\textit{\textit{MPS\_Brick}}), tree tensor network (\textit{\textit{TTN}}), \textit{\textit{TwoLocalRYRZ}} with linear entanglement topology, \textit{\textit{EfficientSU2}} with full entanglement, and \textit{\textit{RealAmplitudes}} with linear entanglement. These ansatzes span a broad spectrum of structural inductive biases, ranging from dense, hardware-efficient circuits to highly structured tensor-network-inspired architectures. The structure of all these ansatz families are shown in Fig.\ref{fig:ansatzes}. 
For each ansatz family, we construct PQCs of the form defined in Eq.~\ref{eq:PQC}, varying the number of qubits $n \in \{2,4,6,8,10\}$ and the number of ansatz repetitions (logical depth) $L \in \{2,4,6,8,10\}$.

\subsection{Hardware-Aware Compilation}
\label{subsec:method_transpile}

Each logical circuit is mapped to a hardware-constrained circuit via hardware-aware compilation:
\begin{equation}
\mathcal{T}_{\mathrm{opt}}: U_{\mathrm{logical}}(\boldsymbol{\theta}) \rightarrow U_{\mathrm{transpiled}}(\boldsymbol{\theta})
\end{equation}

where $\mathcal{T}_{\mathrm{opt}}$ denotes the transpilation procedure under a given transpiler's optimization level. While the resulting unitary $U_{\mathrm{transpiled}}(\boldsymbol{\theta})$ remains functionally equivalent to $U_{\mathrm{logical}}(\boldsymbol{\theta})$, it may differ significantly in circuit structure, depth, and gate composition after transpilation.
The logical circuit prepares the quantum state as in Eq.~\ref{eq:log_state} for \(n\) qubits at logical design level. The corresponding quantum state of transpiled circuit is given by:
\begin{equation}
|\psi_{\mathrm{transpiled}}(\boldsymbol{\theta})\rangle = U_{\mathrm{transpiled}}(\boldsymbol{\theta}) |0\rangle^{\otimes m}
\end{equation}
where $m \geq n$ accounts for possible qubit expansion during layout and routing.


\paragraph{Quantum Backend and Transpilation Pipeline}
Transpiled circuits are obtained using Qiskit's transpiler with the \texttt{IBM FakeBrooklynV2} backend (65 qubits, heavy-hex connectivity). To ensure deterministic compilation, we fix the transpiler seed. The Transpilation pipeline includes:

\begin{itemize}
    \item \textbf{\textit{Basis translation:}} Decomposition of gates into the native set $\{\mathrm{CX}, \mathrm{RZ}, \mathrm{SX}, \mathrm{X}\}$.
    
    \item \textbf{\textit{Layout selection and routing:}} Hardware-aware qubit mapping and routing using the SABRE algorithm~\cite{qiskit_transpilation_sabre}.
    
    \item \textbf{\textit{Optimization:}} Transpilation under all optimization levels ($\mathrm{opt} \in \{0,1,2,3\}$) to study the effect of transpiler aggressiveness.
\end{itemize}

To enable fair comparison, we apply a post-processing step (\texttt{compact\_qubits}) to remove idle qubits and reindex the circuit to the minimal active set.
Transpilation preserves the parameterization of the circuit. Qiskit propagates \texttt{ParameterVector} objects through all compilation stages, ensuring that the transpiled circuit retains the same set and number of trainable parameters as the logical circuit, i.e., $|\boldsymbol{\phi}| = |\boldsymbol{\theta}| = P$, where \(\boldsymbol{\theta}\) and \(\boldsymbol{\phi}\) denote the parameter vectors of the logical and transpiled circuits, respectively.


\paragraph{Choice of a Single Fake Backend}
To isolate the effects of transpilation from logical-level circuit design, we restrict our analysis to \texttt{IBM FakeBrooklynV2}. We use a \textit{fake} backend because active quantum devices introduce noise and calibration drift, which can obscure the structural modifications induced purely by transpilation. Fake backends preserve realistic hardware constraints (e.g., coupling maps and native gate sets) while providing deterministic and reproducible circuit transformations. This enables us to attribute observed changes in expressibility and trainability solely to hardware-aware compilation.
Moreover, we consider a single backend because IBM Quantum devices largely share the heavy-hex topology~\cite{IBM_heavyhex}, resulting in broadly similar routing characteristics across devices. Although different hardware backends may produce slightly different transpiled circuits due to variations in connectivity and native gate sets, our objective is not to perform a cross-device comparison. Rather, we aim to quantify how and to what extent transpilation alone alters expressibility and trainability under a controlled hardware setting.
\subsection{Conversion to PennyLane}
\label{subsec:pennylane_conversion}

We use Qiskit for logical circuit construction and hardware-aware transpilation, and PennyLane for differentiable quantum simulation and metric evaluation. The logical and transpiled circuits are converted to PennyLane-compatible templates using \texttt{qml.from\_qiskit()}.
All simulations are performed using PennyLane’s \texttt{default.qubit} statevector simulator. Separate quantum devices are instantiated for logical and transpiled circuits using \(n_{\mathrm{logical}}\) and \(n_{\mathrm{transpiled}}\) wires, respectively, to account for possible qubit expansion during transpilation.
For each circuit, two QNodes are defined: \textit{(i)} a state QNode returning the full quantum state \( |\psi(\boldsymbol{\theta})\rangle \) for fidelity-based expressibility evaluation, and (ii) an expectation-value QNode returning \( \langle Z_0 \rangle \) for parameter-shift-based gradient computation in the trainability analysis.

\subsection{Quantifying Expressibility Overhead}
\label{subsec:method_expressibility}

We quantify the expressibility of a PQC using the fidelity-distribution-based approach proposed in~\cite{Sim_2019}. For a parameterized circuit \(U(\boldsymbol{\theta})\), the ensemble of output states is defined as:
\begin{equation}
\mathcal{E} =
\left\{
|\psi(\boldsymbol{\theta})\rangle =
U(\boldsymbol{\theta}) |0\rangle^{\otimes n}
:\;
\boldsymbol{\theta} \sim \mathrm{Uniform}[0,2\pi)^P
\right\}
\end{equation}
where \(P\) denotes the number of trainable parameters.
Expressibility is evaluated by comparing the distribution of pairwise state fidelities (Eq.~\ref{eq:fidelity}) against the Haar reference distribution (Eq.~\ref{eq:haar}). The deviation between the two distributions is quantified using the KL divergence:
\begin{equation}
\mathcal{E}_{\mathrm{KL}} =
\sum_{b}
\hat{p}_b
\log
\frac{\hat{p}_b}{\hat{q}_b}
\end{equation}
where \(\hat{p}_b\) and \(\hat{q}_b\) denote the normalized histogram frequencies of the ansatz-induced and Haar distributions, respectively. A small constant \(\epsilon = 10^{-12}\) is added before normalization for numerical stability.
Lower values of \(\mathcal{E}_{\mathrm{KL}}\) indicate higher expressibility, as the circuit-induced distribution more closely matches the Haar distribution.
To evaluate the impact of hardware-aware compilation, we compute expressibility for both logical and transpiled circuits and define the expressibility overhead as in Eq.~\ref{eq:expr_overhead}. A positive overhead indicates reduced expressibility after transpilation, while a negative value indicates improved expressibility.


\subsection{Trainability}
\label{sec:method_trainability}

We consider the cost function of the form as in Eq.~\ref{eq:Cost_F_trainability} and compute gradients using the parameter-shift rule:
\begin{equation}
\frac{\partial C}{\partial \theta_k}
=
\frac{1}{2}
\left[
C\!\left(\boldsymbol{\theta} + \frac{\pi}{2}\mathbf{e}_k\right)
-
C\!\left(\boldsymbol{\theta} - \frac{\pi}{2}\mathbf{e}_k\right)
\right]
\end{equation}
where \(\mathbf{e}_k\) is the \(k\)-th standard basis vector. Gradients are evaluated using PennyLane’s automatic differentiation framework~\cite{bergholm2018pennylane}.
Trainability is then quantified using the mean per-parameter gradient variance:
\begin{equation}
\mathrm{Var}_{\theta}(\nabla C)
=
\frac{1}{P}
\sum_{k=1}^{P}
\mathrm{Var}_{\boldsymbol{\theta}}
\left(
\frac{\partial C}{\partial \theta_k}
\right)
\end{equation}
where \(P\) denotes the number of trainable parameters.
In practice, the variance is estimated empirically using \(N_{\mathrm{grad}}\) randomly sampled parameter vectors:
\begin{equation}
\widehat{\mathrm{Var}}
=
\frac{1}{P}
\sum_{k=1}^{P}
\left[
\frac{1}{N_{\mathrm{grad}}-1}
\sum_{s=1}^{N_{\mathrm{grad}}}
\left(
g_k^{(s)} - \bar{g}_k
\right)^2
\right]
\end{equation}
where
\(
g_k^{(s)}
=
\left.
\frac{\partial C}{\partial \theta_k}
\right|_{\boldsymbol{\theta}=\boldsymbol{\theta}^{(s)}}
\)
and \(\bar{g}_k\) is the sample mean.
Low gradient variance indicates poor trainability and is commonly associated with barren plateau behavior.
To assess the impact of hardware-aware compilation, we compute the gradient variance for both logical and transpiled circuits and define the trainability overhead as in Eq.~\ref{eq:train_overhead}. A positive overhead indicates improved trainability after transpilation, while a negative value indicates degradation.



Both expressibility and trainability estimatation are stochastic due to random parameter sampling. To ensure statistical reliability, both expressibility and trainability are computed and averaged across $M=5$ independent runs.


\section{Results and Discussion}\label{sec:results}


\subsection{Transpilation-Induced Qubit Overhead}
\label{subsec:trans_induced_qubit_overhead}

\begin{figure}[h]
    \centering
    \includegraphics[width=1.0\linewidth]{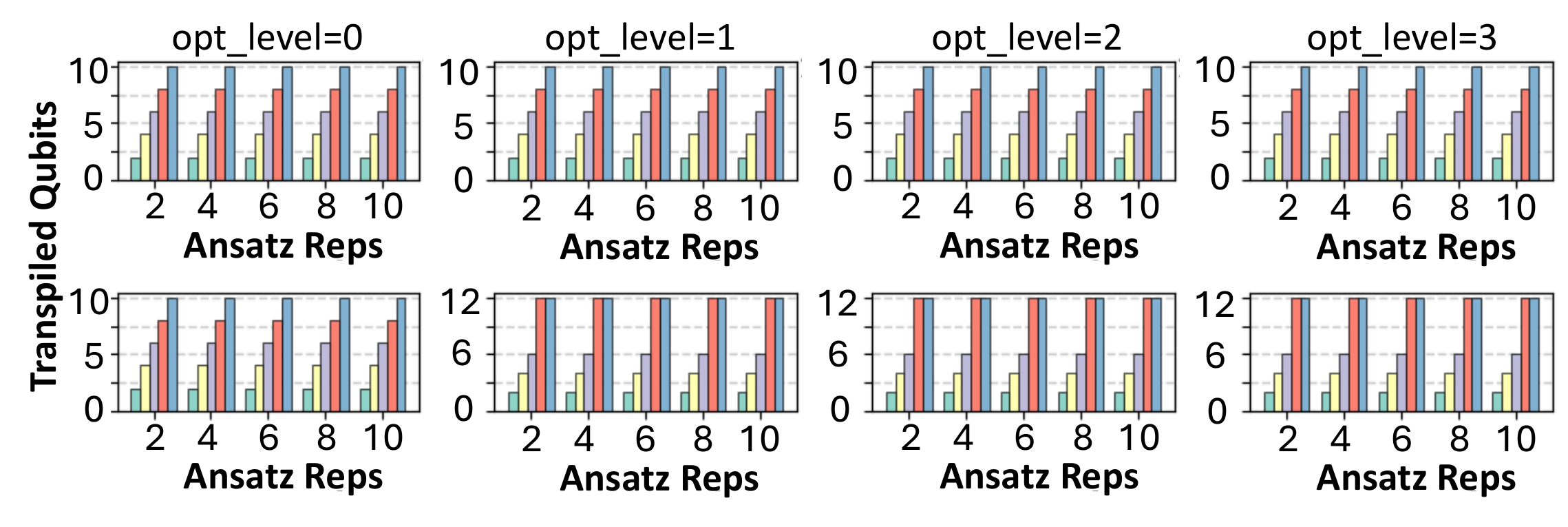}
    \caption{Logical vs. transpiled qubit across all ansatz families. $LQ=$ Logical qubits (designer's intent). Lower pannel is for \textit{HEA\_Ring} ansatz, whereas upper pannel is for all the other ansatz families used in this paper. Qubit count remains same after transpilation  for all ansatz types, across all transpiler opt levels except for HEA\_Ring Ansatz qubit count increases after transpilation for higher qubit count at logical level (8,10) for transpiler opt level of 1-3 except opt level 0.}
    \label{fig:qubit_overhead}
\end{figure}

Fig.~\ref{fig:qubit_overhead} compares the number of qubits at logical design level (LQ) specified by the designer and the number of qubits required after transpilation across tranpsiler optimization levels $0–3$ for all ansatz families.
The upper panel of Fig.~\ref{fig:qubit_overhead} corresponds to \textit{EfficientSU2}, \textit{TTN}, \textit{RealAmplitudes}, \textit{MPS\_Brick}, and \textit{TwoLocalRYRZ}. For these ansatzes, the transpiled qubit count remains identical to the qubit count at logical level across all optimization levels and circuit depths, indicating that transpilation does not introduce any qubit overhead. This suggests that these circuit structures can be mapped onto the target hardware without requiring auxiliary qubits or expanded layouts.
The lower panel of Fig.~\ref{fig:qubit_overhead} shows the results for \textit{HEA\_Ring}, which exhibits slight qubit expansion at larger system sizes. Specifically, for \( \mathrm{LQ}=8 \) and \( \mathrm{LQ}=10 \), the transpiled circuits require $12$ physical qubits for optimization levels $1–3$. This behavior suggests that the transpiler selects alternative layouts involving additional hardware qubits to facilitate routing of the ring entanglement pattern under hardware connectivity constraints.
Notably, this qubit expansion is independent of ansatz repetitions, indicating that the overhead is primarily determined by the interaction between circuit topology and hardware connectivity rather than circuit depth.

\subsection{Transpilation-Induced Depth Overhead}
\begin{figure*}
    \centering
    \includegraphics[width=1.0\linewidth]{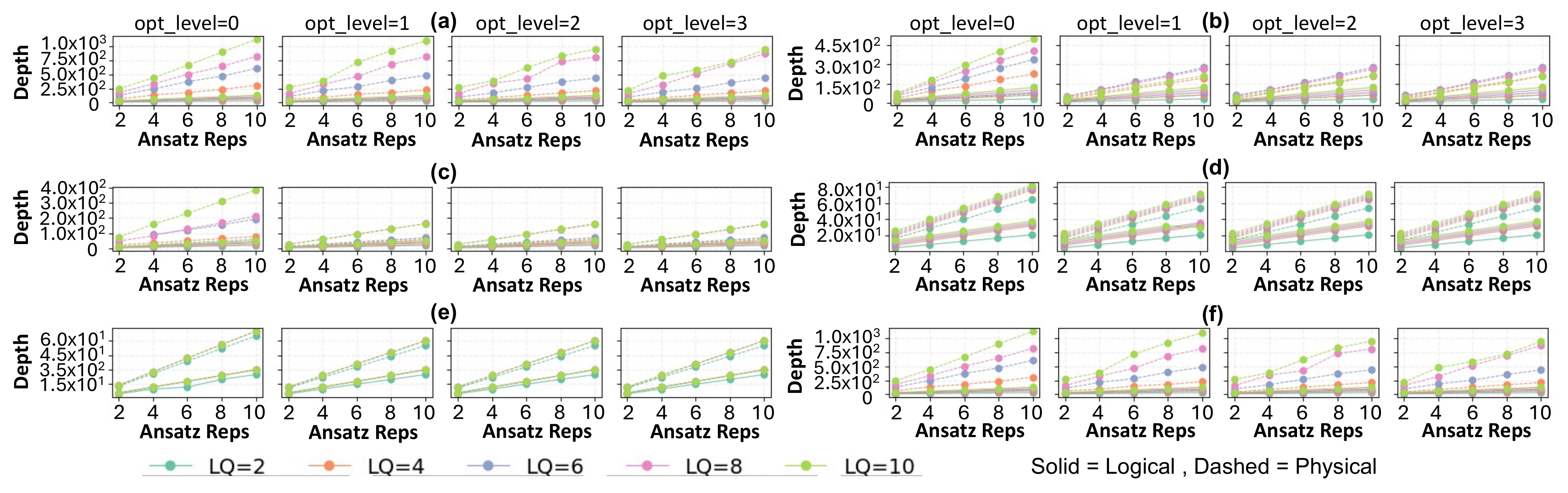}
    \caption{Logical vs. transpiled circuit depth as a function of ansatz repetitions (at logical level) for different logical qubit counts ($LQ = 2--10$). (a) \textit{EfficientSU2}, (b) \textit{HEA\_Ring}, (c) \textit{TTN}\_Tree, (d) \textit{RealAmplitudes}, (e) \textit{MPS\_Brick}, and (f) \textit{TwoLocalRYRZ}. The gap between dashed and solid curves represents the depth overhead induced by hardware-aware transpilation.}
    \label{fig:depth_overhead}
\end{figure*}

Fig.~\ref{fig:depth_overhead} presents the circuit depth before and after hardware-aware transpilation (\texttt{opt\_level = 0-3}) for different ansatz families. The gap between logical and transpiled depth represents the transpilation-induced depth overhead.
Across all ansatzes, transpilation introduces a noticeable increase in circuit depth that generally grows with both ansatz repetitions and the number of qubits at logical design level. This overhead primarily arises from hardware constraints such as limited qubit connectivity, which require additional routing operations, SWAP insertions, and gate decompositions.
A key observation is that the magnitude of the overhead strongly depends on the transpiler optimization level. Lower optimization levels (\texttt{opt\_level=0}) typically produce larger depth overhead due to minimal circuit simplification, whereas higher optimization levels (\texttt{opt\_level=2,3}) partially reduce depth through gate cancellation, commutation analysis, and improved qubit routing. However, this reduction is highly ansatz-dependent.

\textit{EfficientSU2} exhibits the largest depth overhead across all optimization levels, particularly at higher qubit counts (Fig.~\ref{fig:depth_overhead}(a)). Its dense entanglement structure requires extensive routing on hardware with limited connectivity, resulting in significantly deeper transpiled circuits even under aggressive optimization (\texttt{opt\_level=3}) .
\textit{HEA\_Ring} shows comparatively lower depth overhead (Fig.~\ref{fig:depth_overhead}(b)), as its local ring entanglement structure aligns more naturally with hardware connectivity, enabling more effective transpiler optimizations.
\textit{TTN} and \textit{MPS\_Brick} exhibit consistently low overhead across all optimization levels (Fig.~\ref{fig:depth_overhead}(c) and (e), respectively). Their structured and locality-preserving architectures minimize routing requirements and therefore remain relatively insensitive to transpilation.
\textit{RealAmplitudes} and \textit{TwoLocalRYRZ} demonstrate intermediate behavior (Fig.~\ref{fig:depth_overhead}(d) and (f), respectively). Although noticeable overhead appears at lower optimization levels, higher optimization levels reduce the gap more effectively than in densely entangled ansatzes. Nevertheless, the transpiled depth still increases with system size.
These results show that transpilation overhead is jointly determined by ansatz structure and transpiler optimization level. While higher optimization levels can mitigate depth growth, they do not eliminate it, particularly for highly entangled circuits. This highlights that transpilation should be viewed not merely as a resource-level transformation, but as an architecture-dependent modification that can directly influence key PQC properties such as expressibility and trainability.


\subsection{Transpilation-induced Expressibility Overhead}
\begin{figure*}
    \centering
    \includegraphics[width=1.0\linewidth]{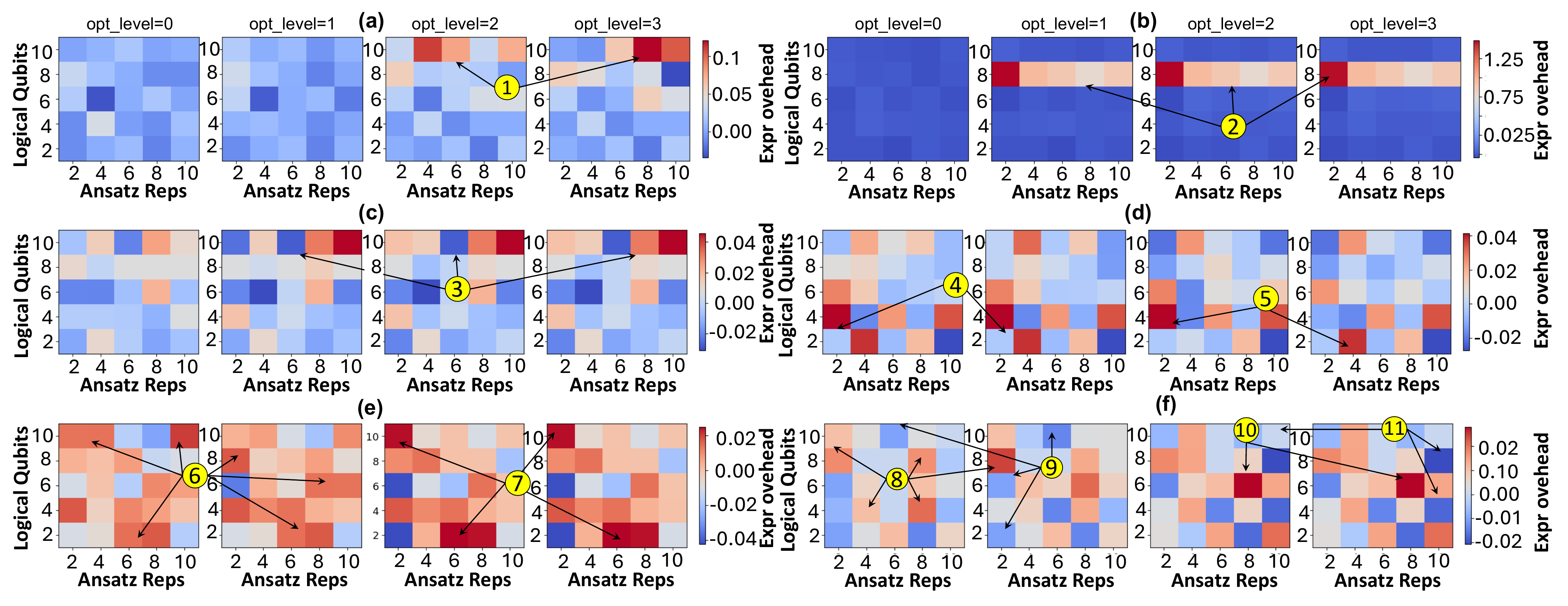}
    \caption{Expressibility overhead (transpiled-logical) of logical and transpiled ansatzes across all ansatz families. Positive values indicate higher KL values of transpiled ansatz (reduced expressibility) and vice versa. Opt\_level denotes transpiler optimization levels from basic (0) to agressive (3). (a) \textit{EfficientSU2} Ansatz, (b) HEA$\_$Ring Ansatz, (c) \textit{TTN} ansatz, (d) Real Amplitudes, (e) \textit{MPS\_Brick}, (f) \textit{TwoLocalRYRZ} }
    \label{fig:expr_overhead}
\end{figure*}

The results of expressibility overhead are presented in Fig.~\ref{fig:expr_overhead}, showing the difference between the KL divergence of transpiled and logical circuits, across all ansatz families, qubit counts, circuit depths, and transpiler optimization levels (\texttt{opt\_level=0-3}). Positive values indicate reduced expressibility after transpilation, while negative values indicate improved expressibility.
Overall, the magnitude and behavior of the overhead strongly depend on the ansatz structure.

\paragraph{Expressibility Overhead in \textit{EfficientSU2}}
\textit{EfficientSU2} exhibits minimal expressibility overhead, with variations largely confined to \([0,0.1]\) across all optimization levels (Fig.~\ref{fig:expr_overhead}(a)). Only small positive deviations appear at larger qubit counts and ansatz repetitions (pointer~\rpoint{1} in (Fig.~\ref{fig:expr_overhead}(a)). Overall, transpilation has limited impact on expressibility, likely because the ansatz is already highly expressive due to its dense entanglement structure.



\paragraph{Expressibility Overhead in \textit{HEA\_Ring}}
\textit{HEA\_Ring} exhibits the largest expressibility overhead among all ansatzes, reaching values up to \(\sim1.25\), as shown in Fig.~\ref{fig:expr_overhead}(b). The degradation is highly localized, appearing primarily at larger qubit counts, and only when transpiler optimization is applied (pointer~\rpoint{2} in (Fig.~\ref{fig:expr_overhead}(b)). Notably, this coincides with the qubit-expansion region observed in Section~\ref{subsec:trans_induced_qubit_overhead}, indicating that the additional routing qubits do not improve expressibility.

\paragraph{Expressibility Overhead in \textit{TTN}}
\textit{TTN} ansatz exhibits relatively small expressibility overhead across all transpiler optimization levels, with variations confined to \([-0.02,0.04]\), as shown in Fig.~\ref{fig:expr_overhead}(c). The overhead remains centered near zero, indicating limited sensitivity to transpilation. Higher optimization levels introduce slightly more structured positive deviations at larger qubit counts and ansatz repetitions (pointer~\rpoint{3}, in Fig.~\ref{fig:expr_overhead}(c)), but the overall magnitude remains small. This robustness is likely due to the hierarchical and locality-preserving structure of \textit{TTN} ansatz, which naturally aligns with hardware connectivity and minimizes routing overhead during transpilation.


\paragraph{Expressibility Overhead in \textit{RealAmplitudes}}
The expressibility overhead for \textit{RealAmplitudes} remains within a narrow range of \([-0.020,0.040]\), indicating limited sensitivity to transpilation, as shown in Fig.~\ref{fig:expr_overhead}(d). At lower optimization levels (\texttt{opt\_level=0-1}), zero or slightly deviations dominate most configurations, with only a few localized positive regions appearing at shallow depths and lower qubit counts (pointer~\rpoint{4} in Fig.~\ref{fig:expr_overhead}(d)). As the optimization level increases (\texttt{opt\_level=2-3}), the negative overhead becomes more widespread and consistent, while positive deviations remain sparse (pointer~\rpoint{5} in in Fig.~\ref{fig:expr_overhead}(d)). Overall, transpilation tends to slightly improve the expressibility of \textit{RealAmplitudes}, particularly at higher optimization levels.



\paragraph{Expressibility Overhead in \textit{MPS\_Brick}}
The expressibility overhead for \textit{MPS\_Brick} lies within \([-0.040,0.020]\), as shown in (Fig.~\ref{fig:expr_overhead}(e)). At lower optimization levels (\texttt{opt\_level=0-1}), the overhead is predominantly positive (pointer~\rpoint{6} in Fig.~\ref{fig:expr_overhead}(e)), with only a few localized negative regions. 
As the optimization level increases (\texttt{opt\_level=2-3}), deviations become more pronounced in both directions, with stronger positive deviations (pointer~\rpoint{7} in Fig.~\ref{fig:expr_overhead}(e)), and negative regions emerging across configurations. Overall, \textit{MPS\_Brick} remains relatively stable under transpilation, although higher optimization levels introduce stronger configuration-dependent variations. The larger magnitude of negative deviations suggests that transpilation-induced improvements in expressibility can occasionally outweigh degradations.
\paragraph{Expressibility Overhead in \textit{TwoLocalRYRZ}}
The expressibility overhead for \textit{TwoLocalRYRZ} is in range \([-0.02,0.02]\), as shown in (Fig.~\ref{fig:expr_overhead}(f)). At lower optimization levels (\texttt{opt\_level=0-1}), the overhead exhibits mixed positive and negative deviations across configurations (pointers~\rpoint{8} and \rpoint{9} in Fig.~\ref{fig:expr_overhead}(f)). As the optimization level increases (\texttt{opt\_level=2-3}), the magnitude of deviations becomes more pronounced in both directions (pointers~\rpoint{10} and \rpoint{11} in Fig.~\ref{fig:expr_overhead}(f)), producing a more contrasted pattern. Higher optimization levels amplify configuration-dependent expressibility variations without introducing a consistent directional trend.



\paragraph{Comparative Analysis of Expressibility Overhead Across All Ansatz Families}
Across all ansatz families, expressibility overhead is strongly influenced by ansatz structure and transpilation-induced circuit modifications. \textit{HEA\_Ring} exhibits the largest degradation, closely aligned with regions of increased physical qubit usage. In contrast, \textit{EfficientSU2} remains largely unaffected despite significant depth overhead, suggesting that already highly expressive circuits are less sensitive to structural perturbations.
Structured ansatzes such as \textit{TTN\_Tree} and \textit{MPS\_Brick} demonstrate strong robustness due to their locality-preserving architectures, which minimize routing overhead during transpilation. \textit{RealAmplitudes} and \textit{TwoLocalRYRZ} exhibit moderate, configuration-dependent variations, with higher optimization levels generally amplifying expressibility changes without introducing a consistent directional trend.
Overall, these results show that expressibility overhead is governed not only by circuit size or depth, but also by how hardware-aware compilation reshapes circuit structure.

\begin{figure*}
    \centering
    \includegraphics[width=1.0\linewidth]{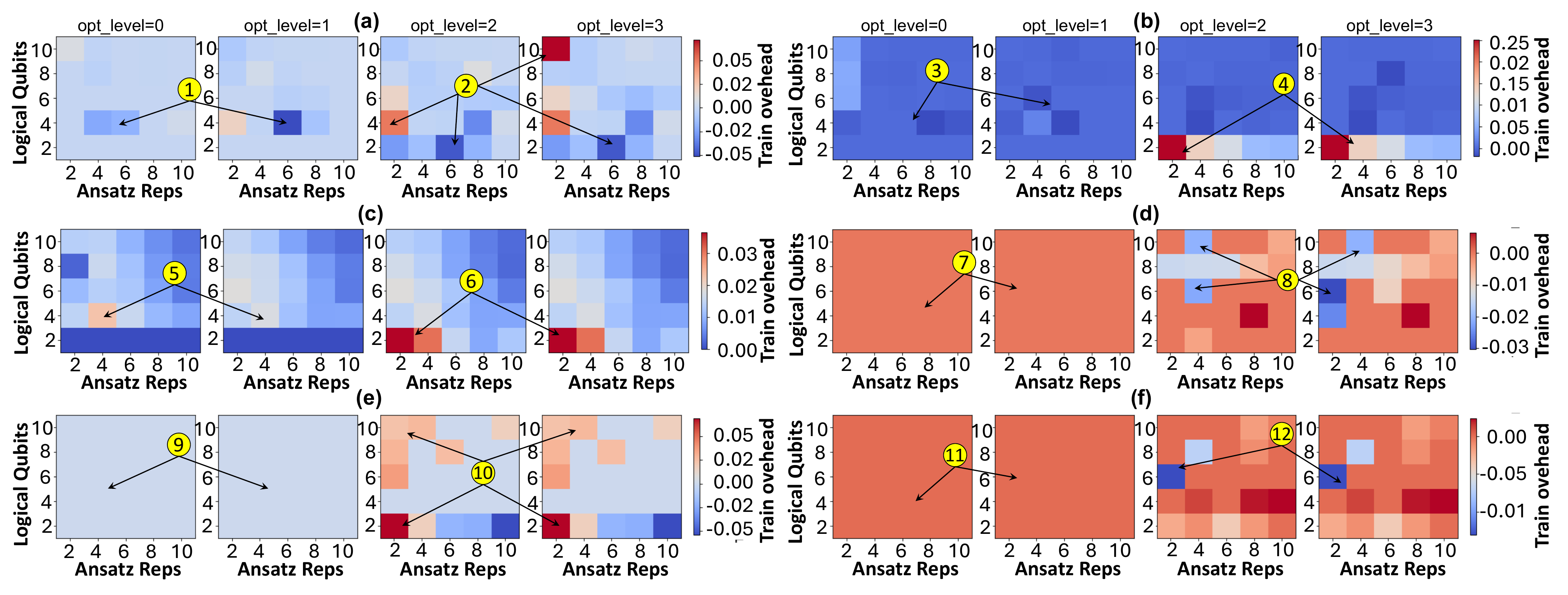}
    \caption{Trainability overhead (transpiled-logical) of logical and transpiled ansatzes across all ansatz families. Positive values indicate higher trainability of transpiled ansatz and vice versa. Opt\_level denotes transpiler optimization levels from basic (0) to agressive (3). (a) \textit{EfficientSU2} Ansatz, (b) HEA$\_$Ring Ansatz, (c) \textit{TTN} ansatz, (d) Real Amplitudes, (e) \textit{MPS\_Brick}, (f) \textit{TwoLocalRYRZ} }
    \label{fig:train_overhead}
\end{figure*}


\subsection{Transpilation-Induced Trainability Overhead}

Fig.~\ref{fig:train_overhead} presents the trainability overhead (transpiled - logical) across different ansatz families, qubit counts, circuit depths, and transpiler optimization levels (\texttt{opt\_level=0-3}). Positive values indicate improved trainability after transpilation, while negative values indicate degraded trainability.

\paragraph{Trainability Overhead in \textit{EfficientSU2}}
The trainability overhead for \textit{EfficientSU2} is in range \([-0.004,0.004]\), with most configurations centered near zero (Fig.~\ref{fig:train_overhead}(a)). At lower optimization levels (\texttt{opt\_level=0-1}), only small localized negative deviations are observed (pointer~\rpoint{1} in Fig.~\ref{fig:train_overhead}(a)), while higher optimization levels (\texttt{opt\_level=2-3}) produce slightly more pronounced positive and negative regions (pointer~\rpoint{2} in Fig.~\ref{fig:train_overhead}(a)). Overall, \textit{EfficientSU2} exhibits relatively stable trainability under transpilation, although the effect remains configuration-dependent.


\paragraph{Trainability Overhead in \textit{HEA\_Ring}}
The trainability overhead for \textit{HEA\_Ring} is in range \([0,0.025]\), and is only positive or zero (Fig.~\ref{fig:train_overhead}(b)), indicating that transpilation generally preserves or slightly improves trainability. At lower optimization levels (\texttt{opt\_level=0-1}), the overhead remains close to zero across most qubit counts and ansatz repetitions (pointer~\rpoint{3} in Fig.~\ref{fig:train_overhead}(b)). At higher optimization levels (\texttt{opt\_level=2-3}), a localized region of strong positive overhead emerges at small qubit counts (\(LQ=2\)) (pointer~\rpoint{4} in Fig.~\ref{fig:train_overhead}(b)), while the remaining configurations continue to exhibit only minor variations. Overall, \textit{HEA\_Ring} also maintains relatively stable trainability under transpilation, with improvements confined to specific low-qubit regimes.

\paragraph{Trainability Overhead in \textit{TTN}}
The trainability overhead for \textit{TTN} remains within \([0,0.03]\), with predominantly positive values across most configurations (Fig.~\ref{fig:train_overhead}(c)), indicating that transpilation generally preserves or improves trainability. At lower optimization levels (\texttt{opt\_level=0-1}), the overhead is smooth across qubit counts and ansatz repetitions, indicating no impact of transpilation (pointer~\rpoint{5} in Fig.~\ref{fig:train_overhead}(c)). At higher optimization levels (\texttt{opt\_level=2-3}), more structured positive regions emerge, particularly at lower qubit counts and relatively higher depths, with the magnitude gradually decreasing as the number of qubits increases (pointer~\rpoint{6} in Fig.~\ref{fig:train_overhead}(c)). Overall, \textit{TTN\_Tree} exhibits consistently improved trainability or no changes under transpilation.


\paragraph{Trainability Overhead in \textit{RealAmplitudes}}
The trainability overhead for \textit{RealAmplitudes} lies within \([-0.03,0.00]\), indicating that transpilation generally tends to reduce trainability (Fig.~\ref{fig:train_overhead}(d)). At lower optimization levels (\texttt{opt\_level=0-1}), the overhead remains close to zero across most configurations (pointer~\rpoint{7} in Fig.~\ref{fig:train_overhead}(d)), suggesting strong robustness to transpilation-induced perturbations. At higher optimization levels (\texttt{opt\_level=2-3}), negative overhead becomes more pronounced across multiple configurations (pointer~\rpoint{8} in Fig.~\ref{fig:train_overhead}(d)), indicating suppressed gradient variance and degraded trainability under more aggressive transpiler optimizations. Overall, \textit{RealAmplitudes} maintains stable trainability at lower optimization levels, but exhibits moderate degradation at higher optimization levels.



\paragraph{Trainability Overhead in \textit{MPS\_Brick}}
The trainability overhead for \textit{MPS\_Brick} remains within a narrow range of \([-0.005,0.005]\), indicating limited sensitivity to transpilation (Fig.~\ref{fig:train_overhead}(e)). At lower optimization levels (\texttt{opt\_level=0-1}), the overhead remains nearly zero across all configurations (pointer~\rpoint{9} in Fig.~\ref{fig:train_overhead}(e)), demonstrating strong robustness to transpilation. At higher optimization levels (\texttt{opt\_level=2-3}), more structured positive and negative deviations emerge, particularly at lower qubit counts, while higher-qubit configurations exhibit only minor changes (pointer~\rpoint{10} in Fig.~\ref{fig:train_overhead}(e)). Overall, \textit{MPS\_Brick} maintains highly stable trainability under transpilation, with only small perturbations appearing under aggressive optimization.



\paragraph{Trainability Overhead in \textit{TwoLocalRYRZ}}
The trainability overhead for \textit{TwoLocalRYRZ} remains within a narrow range of \([-0.01,0.01]\), indicating extremly limited sensitivity to transpilation (Fig.~\ref{fig:train_overhead}(f)). At lower optimization levels (\texttt{opt\_level=0-1}), the overhead remains close to zero across all configurations (pointer~\rpoint{11} in Fig.~\ref{fig:train_overhead}(f)), suggesting negligible impact on gradient variance. At higher optimization levels (\texttt{opt\_level=2-3}), a few localized negative deviations emerge, particularly at intermediate qubit counts and lower ansatz repetitions (pointer~\rpoint{12} in Fig.~\ref{fig:train_overhead}(f)), indicating slight degradation in trainability under more aggressive optimization. Overall, \textit{TwoLocalRYRZ} remains largely robust to transpilation-induced trainability changes.


\paragraph{Comparative Analysis of Trainability Overhead Across All Ansatz Families}
Across all ansatz families, trainability overhead exhibits a strong dependence on circuit structure and transpiler optimization level. \textit{TTN} and \textit{HEA\_Ring} generally show positive overhead in specific configurations, indicating localized improvements in gradient variance, particularly at lower qubit counts and higher optimization levels. In contrast, \textit{EfficientSU2} remains largely stable, with only small positive and negative deviations across configurations.
\textit{MPS\_Brick} and \textit{TwoLocalRYRZ} demonstrate strong robustness, exhibiting near-zero overhead at lower optimization levels and only minor variations under more aggressive optimization. In contrast, \textit{RealAmplitudes} shows a consistent negative bias, indicating moderate trainability degradation at higher optimization levels.
Overall, trainability overhead remains relatively small across all ansatz families compared to expressibility overhead, although clear configuration-dependent patterns emerge depending on ansatz structure and transpiler optimization strength.


\subsection{Expressibility–Trainability Tradeoff under Transpilation}

In VQAs, it is commonly understood that a trade-off exists between expressibility and trainability. Highly expressive ansatzes are capable of exploring a larger portion of the Hilbert space, but this increased expressibility is often associated with the emergence of barren plateaus, leading to poor trainability. 
However, from a hardware-aware compilation perspective, which is a mandatory step when deployment on actual hardware is required, our results show that this trade-off does not hold uniformly. Across different ansatz families, transpilation modifies expressibility and trainability in a non-uniform and often decoupled manner. In several cases, changes in expressibility are not accompanied by corresponding changes in trainability. For example, \textit{\textit{EfficientSU2}} remains largely stable in both metrics despite significant depth overhead, while \textit{\textit{RealAmplitudes}} shows reduced trainability without a consistent improvement in expressibility. Similarly, structured ansatzes such as \textit{\textit{TTN}\_Tree} and \textit{\textit{MPS\_Brick}} maintain stable trainability with only minor variations in expressibility.
Overall, these observations suggest that the expressibility–trainability trade-off, as understood at the logical circuit level, does not directly translate to hardware-aware implementations. Instead, transpilation introduces architecture- and configuration-dependent effects that can decouple these two properties, highlighting the need to evaluate both metrics directly on transpiled hardware-deployable circuits.


\section{Conclusion}
We presented a hardware-aware analysis of expressibility and trainability in variational quantum algorithms by systematically comparing logical circuits with their transpiled counterparts. Our results show that transpilation is not merely a resource optimization step, but can fundamentally alter both the representational capacity and the optimization landscape of parameterized quantum circuits.
We further demonstrate that these effects are strongly ansatz-dependent. In particular, expressibility and trainability can change independently under transpilation, indicating that the commonly assumed trade-off between them does not consistently hold in hardware-aware setting.
Overall, our findings highlight that logical-level analysis is not always reliable predictors of hardware-level behavior. This underscores the need for hardware-aware evaluation of PQCs, as conclusions drawn at the abstract design level may not directly translate to practical, hardware-constrained implementations.

\section*{Acknowledgment}
This work was supported in part by the NYUAD Center for Quantum and Topological Systems (CQTS), funded by Tamkeen under the NYUAD Research Institute grant CG008.

\end{spacing}

\begin{spacing}{0.99}
    
\bibliographystyle{IEEEtran}

\bibliography{refs}
\end{spacing}

\end{document}